\documentclass[a4paper]{article} 
\usepackage{epsfig}
\usepackage{graphics}
\usepackage{latexsym}
\usepackage{amsmath}
\usepackage{amssymb}
\usepackage{rotating}
\usepackage{subfigure}
\usepackage{bm}
\usepackage{color}
\usepackage{blindtext}
\usepackage{multirow}
\usepackage{hyperref}

\title{ Three-Body $B^0_{(s)}$ to $\phi \pi^+ \pi^-$ Decays
}
\author{T. Estabar\footnote{T.estabar@semnan.ac.ir},
H. Mehraban\footnote{hmehraban@semnan.ac.ir}\\
Physics Department, Semnan University\\
P.O.Box 35195-363, Semnan, Iran}
\begin{document}
\maketitle
\normalsize
\begin{abstract}

We interest to investigate of the three-body decays of $B^{0}$ meson to $ \phi\pi^+\pi^- $ and  $B_s^{0}$ meson to $ \phi  \pi^{+}\pi^{-}$. Hadronic three-body decays include both nonresonant and resonant contributions, on the basis of the factorization hypothesis. In this analysis, resonant structure is exhibited only in the $ \pi^+\pi^-$ channel which the resonant contribution can be described by S-wave, P-wave and D-wave $ \pi^+\pi^-$ contribution from $ f_0(980)$, $ \rho$ and $f_2$ mesons and other possible resonance. Therefore, the theoretical values at the scale $ m_b$ are $(1.69 \pm 0.19  )\times10^{-7}$ and $(3.28\pm 0.17 )\times10^{-6}$, while the experimental results of them are $(1.82 \pm 0.25 )\times10^{-7}$ and $(3.48\pm 0.23 )\times10^{-6}$, respectively. Comparing computation analysis values with experimental values show that our results at the scale $\mu_b $ are in agreement with them.

\end{abstract}

\section{Introduction}
\normalsize
The three-body decays of $B^{0}_s \longrightarrow \phi \pi^+\pi^+$ and $B^{0} \longrightarrow \phi \pi^+\pi^-$ have not been observed before and were recorded by the LHCb experiment and tabulated by the Particle Data Group (PDG)~\cite{Aaij:2016qnm, Patrignant:2016cp}.

 Three-body decay of D meson to $ K\pi\pi $ has been analyzed long time ago~\cite{Dalitz:1953cp}. In these sorts of three-body decay, final state mesons are supposed to be light. The momentum of output mesons and matrix elements of amplitude are written by variables $ s=(p_{B}-p_{3})^{2} $  and $ t=(p_{B}-p_{1})^{2} $~\cite{Bajc:1998bs,Fajfer:1998yc}, and Dalitz plot technique should be used integral from $ s_{min}$, $t_{min} $ to $ s_{max}$, $t_{max} $ for computing of the decay width. This double integral encompasses all angles between the momenta of output mesons. The amplitudes of three-body decays can be obtained, the Feynman quark diagrams should be plotted. The direct three-body $B^{0} \rightarrow \phi \pi^+\pi^-$ decay receive two separate parts: one from the point like weak transition and the other from the pole diagrams that involves three points or four points strong vertices. First, we consider parameters which appear in the factorized term of the hadronic matrix element. In the case of $ < B^0 \rightarrow \pi\pi>\times <0 \rightarrow \phi>$, both $\pi$ mesons are located in the form factor. In fact, two-meson matrix element transition of B meson is described to the $\pi$ mesons. The $\phi$ is placed in the decay constant. In addition, there is the emission-annihilation process $ < B^0 \rightarrow 0>\times <0 \rightarrow \phi \pi\pi>$. The total amplitude is computed a sum of amplitudes from nonresonant and resonant contribution. The resonant contribution evaluated in Dalitz plot analysis. The matrix element of amplitude is related to the multiplying the B meson to the pion pair transition in the different waves by the vacuum to the $ \phi$ meson transition. There are several resonances in S-wave, P-wave and D-wave $ \pi^+ \pi^-$ contributions with $ \pi^+ \pi^-$ invariant mass in the range $ 400< m(\pi^+ \pi^)<1600$ Mev/$c^2$. As the analysis of the resonant contribution defined by the Breit-Wigner function is used to investigate the intermediate states $f_0 \phi $, $ \rho \phi $ and $f_2 \phi $. Other three-body $B^{0}_s \rightarrow \phi \pi^+\pi^-$ decay is investigated. The parameters appeared in the factorized term are $ < B^0_s \rightarrow \phi>\times <0 \rightarrow \pi\pi>$ and $ < B^0_s \rightarrow 0>\times <0 \rightarrow \phi \pi\pi>$. The $0 \rightarrow \pi\pi $ matrix element is supposed to be proportional to the pion scalar, vector and tensor form factors, then, the different resonances $ f_{0i}$, $\rho_i$ and $f_2$ show in the $\pi^+\pi^- $ interaction.

\section{The Analysis of Amplitude }
\normalsize

 In this section, the amplitude and branching ratio of $B^0 \rightarrow \phi \pi^+ \pi^-$ and $B^0_s \rightarrow \phi \pi^+ \pi^-$are obtained by using factorization method. We have to contemplate the nonresonant and resonant contributions separately.
 Hadronic weak decays are evaluated by the effective weak Hamiltonian~\cite{Cheng:2015ckx},\\
\begin{eqnarray}
H_{eff} = \dfrac{G_{F}}{\sqrt{2}}\sum_{i}[V_{CKM} C_{i}(\mu) O_{i}(\mu)].
\end{eqnarray}
Here, the $G_{F}$ defines the Fermi coupling constant, the coefficients $V_{CKM}$ are explained elements of Cabibbo-Kobayashi-Maskawa (CKM) \cite{buchalla:1996f,dEnterria:2016rbf}, $C_{i}(\mu)$ are the Wilson coefficients~\cite{buras:1995} and $O_{1,2}$ denote current-current operators, $O_{3-6}$ being penguin operators and $O_{7-10}$ the electroweak penguin operators~\cite{Buras:1994qa}.\\

 \subsection{Nonresonant contribution} In the factorization approach, the Feynman diagrams for three-body $B^0 \rightarrow \phi \pi^+ \pi^-$ decay are depicted in Fig. 1 which including the factorizable and non-factorizable diagrams. In the penguin level, the two $\pi$ mesons are located in form factor and the $\phi$ meson is located in the decay constant. The emission annihilation diagrams in which $\phi$ is emitted via gluons exchange are named hairpin diagrams.The current analysis includes non factorizable effects. The non factorizable terms are dependent on vertex corrections and hard spectator interactions \cite{Cheng:2001ez}. The non factorizable  diagrams depicted in Fig. 2 should be taken into account. For studying $B^0 \rightarrow \phi \pi^+ \pi^-$, we use the QCD factorization framework, which includes theoretical property of QCD like color transparency and hard scattering.Thus, The amplitude of this decay includes $<B^{0} \longrightarrow \pi^+\pi^->\times<0\longrightarrow \phi>$ and $<B^{0} \longrightarrow 0> \times <0\longrightarrow \phi \pi^{+}\pi^{-}>$. Thus, the decay amplitude is given by
    \begin{align}
  \nonumber
 < \phi \pi^+ \pi^- \vert H_{eff} \vert B^0>&=\dfrac{iG_{F}}{2\sqrt{2}}((a_3+a_5+a_7)\lambda_p <\pi^+ \pi^- \vert (\bar{b}d)_{V-A} \vert B^{0}>\\
  \nonumber
  & \times <\phi \vert (\bar{s}s)_{V-A} \vert 0>+(a_{2}V_{ub}V^{*}_{us} +(a_3+a_5+a_7)\lambda_p)\\
    \nonumber
  & \times <0\vert (\bar{b}d)_{V-A} \vert B^{0}> <\phi \pi^{+}\pi^{-}\vert (\bar{u}u)_{V-A} \vert 0> +(a_3\\
  \nonumber
& +a_5+a_7+a_4+a_10)\lambda_p <0\vert (\bar{b}d)_{V-A} \vert B^{0}>\\
\nonumber
& \times <\phi \pi^{+}\pi^{-}\vert (\bar{d}d)_{V-A} \vert 0>+(a_6+a_8)\lambda_p <0\vert (\bar{b}d) \vert B^{0}>\\
& \times <\phi \pi^{+}\pi^{-}\vert (\bar{d}d) \vert 0>,
&\end{align}\\
  \begin{figure}[t]
\centering
\includegraphics[scale=.62]{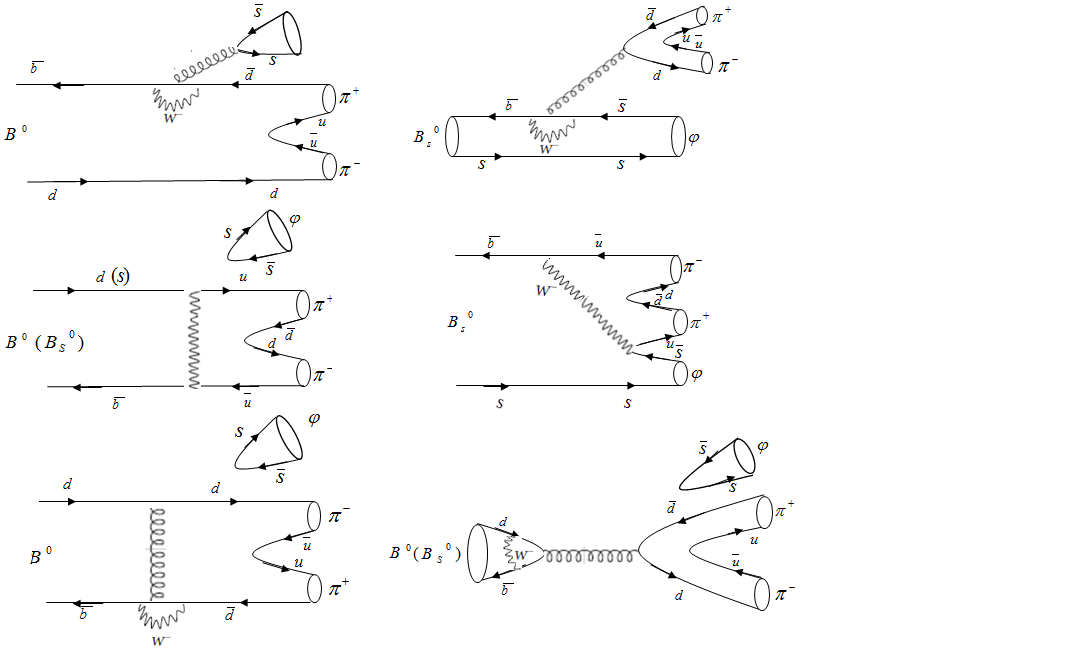}
\caption{The diagrams for $ B^0(B^0_S)\longrightarrow \phi \pi^{+}\pi^{-} $ decay. }
\end{figure}

 \begin{figure}[t]
\centering
\includegraphics[scale=.71]{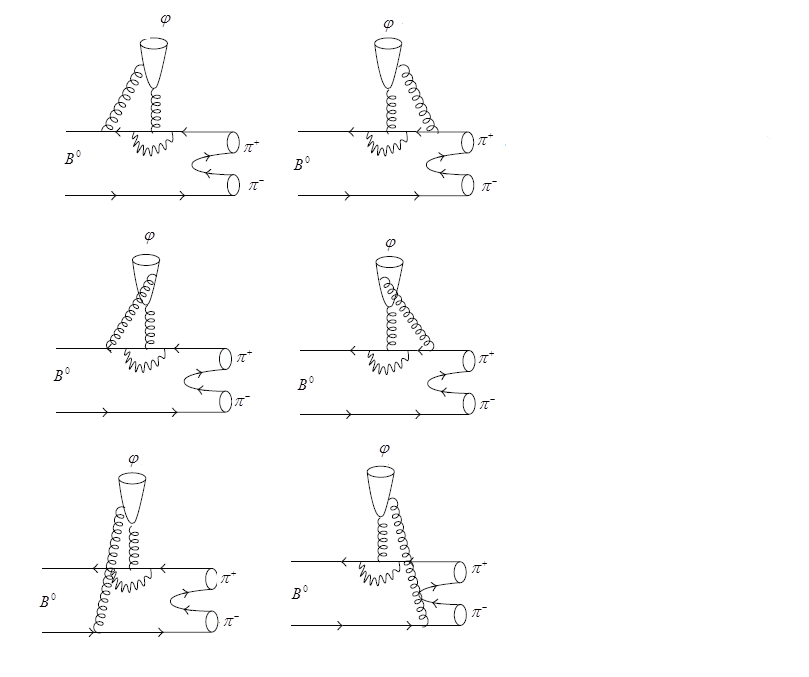}
\caption{Nonfactorizable diagrams for $ B^0\longrightarrow \phi\pi^{+}\pi^{-} $ decay. }
\end{figure}
where $\lambda_{p}=\sum_{p=u,c}V_{pb}V^*_{pd}$. The three-body matrix elements $<\pi^{+}\pi^{-}\vert (\bar{b}d)_{V-A}\vert B^{0}>$  have the following general form as~\cite{Cheng:2013dua}\
\begin{eqnarray}
 <\pi^{-}(p_{1})\pi^{+}(p_{2})\vert (\bar{b}d)_{V-A}\vert ِB^{0}(p_{B})>=ir(p_{B}-p_{1}-p_{2})_{\mu}+iw_{+}(p_{1}+p_{2})_{\mu}\\
 \nonumber
+iw_{-}(p_{1}-p_{2})_{\mu}+h\varepsilon_{\mu\nu\alpha\beta}p_{B}^{\nu}(p_{1}+p_{2})^{\alpha}(p_{2}-p_{1})^{\beta} .\quad\quad
\end{eqnarray}

We need to consider point-like and pole diagrams depicted in Fig. 3. We also require the strong coupling constant of $ B^{\ast}B\pi $ and $ BB\pi \pi $. The form factors $w_{\pm}$ and r for the nonresonant decay are evaluated from these diagrams as~\cite{Cheng:2014uga}\  \\
\begin{eqnarray}\label{r}
\nonumber
r&=&\dfrac{f_B}{2f_\pi^2}-\dfrac{f_B}{f_\pi^2}\dfrac{p_{B}.(p_{2}-p_{1})}{(p_{B}-p_{1}-p_{2})^{2}-m_{B}^{2}}+\dfrac{2gf_{B^*}}{f_{\pi}^{2}}\sqrt{\dfrac{m_{B}}{m_{B^{*}}}}\dfrac{(p_{B}-p_{1}).p_{1}}{(p_{B}-p_{1})^{2}-m_{B^{*}}^{2}}  
\nonumber\\
& -&\dfrac{4g^2f_B}{f_\pi^2} \dfrac{m_B m_{B^*}}{(p_B-p_1-p_2)^2-m_B^2} \dfrac{p_1.p_2-p_1.(p_B-p_1)p_2.(p_B-p_1)/m_{B^*}^2}{(p_B-p_1)^2-m_{B^*}^2},\nonumber
 \end{eqnarray}
\begin{equation}\label{w_+}
w_+=- \dfrac{g}{f_\pi^2} \dfrac{f_{B^{*}}m_{B^{*}}\sqrt{m_{B^{*}}m_{B}}}{(p_{B}-p_{1})^{2}-m_{B^*}^2}(1-\dfrac{(p_B-p_1).p_1}{m_{B^{*}}^2})+\dfrac{f_B}{2f_\pi^2},\nonumber \quad\quad\quad\quad
\end{equation}
\begin{equation}
\nonumber
w_{-}=\dfrac{g}{f_{\pi}^{2}} \dfrac{f_{B^{*}}m_{B^{*}}\sqrt{m_{B^{*}}m_{B}}}{(p_{B}-p_{1})^{2}-m_{B^{*}}^{2}}(1+\dfrac{(p_{B}-p_{1}).p_{1}}{m_{B^{*}}^{2}}),\quad\quad\quad\quad\quad\quad\quad\quad
\end{equation}
\begin{equation}
h=2g^2\dfrac{f_B}{f_{\pi}}\dfrac{m_B^2}{(m_B^2-m_{\phi}^2-s)(t+m_B^2-m_{\pi}^2)}, \quad\quad\quad\quad\quad\quad\quad\quad\quad
\end{equation}

where g is a heavy-flavor independent strong coupling. 
 The decay constants of vector meson are defined~\cite{Wolfgang:2014lms}
\begin{eqnarray}
<0 \vert (\bar{s}s)_{V-A} \vert \phi(p_3,\varepsilon)>=f_{\phi}m_{\phi}\varepsilon_{\mu}^*.
\end{eqnarray}

  \begin{figure}[t]
\centering
\includegraphics[scale=.73]{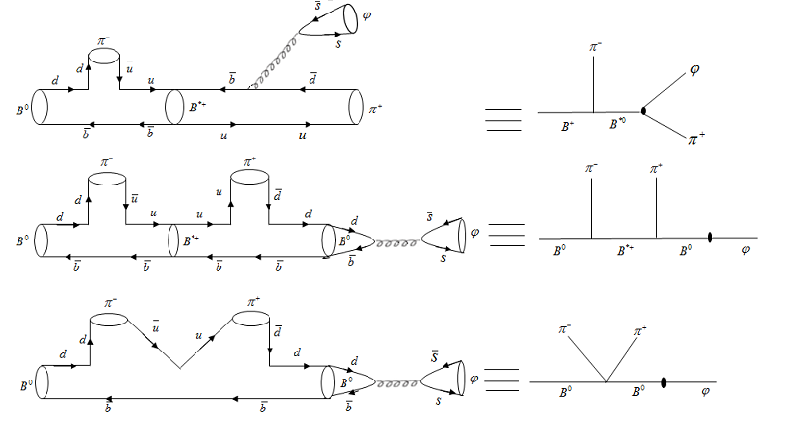}
\caption{Point like and pole diagrams for $ B^0 \longrightarrow \phi \pi^{+}\pi^{-} $ decay. }
\end{figure}
 
For the multiplication of the matrix element, we have
\begin{eqnarray}
\nonumber
 <\pi^{-}(p_{1})\pi^{+}(p_{2})\vert (\bar{b}d)_{V-A}\vert ِB^{0}(p_{B})><0 \vert (\bar{s}s)_{V-A} \vert \phi(p_3,\varepsilon)> =if_{\phi}m_{\phi}(r\varepsilon.p_3\\
 +w^+ \varepsilon.(p_1+p_2)+w^-(p_2-p_1)),\quad\quad \quad
\end{eqnarray}
where under the lorentz condition $ \varepsilon.p_3=0 $. The structure of polarization vector can be described as
\begin{eqnarray}
\nonumber
&\varepsilon^{\lambda=0}=(\vert p_3\vert, 0, 0, p^{0}_3)/m_{3} ,\\
& \varepsilon^{\lambda=\pm 1}=\dfrac{\mp(0, 1, \pm i, 0)}{\sqrt{2}}.\quad\quad
\end{eqnarray}
The energy-momentum conservation could be shown by:
\begin{equation}
p_{B}=p_{1}+p_{2}+p_{3} .
\end{equation}
Define the three following invariants that are not independent as:
\begin{eqnarray}
\nonumber
s_{12}=(p_{1}+p_{2})^{2}=(p_{B}-p_{3})^{2} ,\\
\nonumber
s_{13}=(p_{1}+p_{3})^{2}=(p_{B}-p_{2})^{2} ,\\
s_{23}=(p_{2}+p_{3})^{2}=(p_{B}-p_{1})^{2} ,
\end{eqnarray}
according to definition of 4 momentum conservation, we have from invariants
\begin{equation}
s_{12}+s_{13}+s_{23}=m^2_{B}+m^2_{1}+m^2_{2}+m^2_{3} ,
\end{equation}
we let $s_{12}=s$ and $s_{23}=t$. In the center of mass of $\pi^-(p_1)$ and $\pi^+(p_2)$, we find
\begin{align}
\nonumber
\vert p_1\vert &=\vert p_2\vert=\dfrac{1}{2}\sqrt{s-4m_1^2},\\
\nonumber
p_1^0 &=p_2^0=\dfrac{1}{2}\sqrt{s},\\
\nonumber
\vert p_3\vert &=\dfrac{1}{2\sqrt{s}}\sqrt{(m_B^2-m_3^2-s)^2-4sm_3^2},\\
 p_3^0 &=\dfrac{1}{2\sqrt{s}}(m_B^2-m_3^2-s).
\end{align}

and the cosine of the helicity angle $\theta$ between direction of $ p_2$ and that $ p_3$ reads
\begin{eqnarray}
cos\theta= \dfrac{1}{4\vert p_2\vert\vert p_3\vert}(m_B^2+m_3^2+2m_2^2-s-2t).
\end{eqnarray}
With these definitions, we obtain as
\begin{align}
\nonumber
\varepsilon &.(p_1+p_2)=2p_1^0\varepsilon^0 ,\\
\varepsilon &.(p_2-p_1)=2\vert \varepsilon \vert \vert p_1 \vert cos\theta.
\end{align}

 the matrix elements of annihilation process are written by
\begin{eqnarray}
\nonumber
<\pi^{-}(p_{1})\pi^{+}(p_{2})\phi(p_{3}) \vert (\bar{d}d)_{V-A} \vert 0> =\dfrac{2i}{f_{\pi}}(p_{2\mu}-\dfrac{p_{B}.p_{2}}{p^2_{B}-p^2_{3}}p_{B\mu}) F^{\phi\pi\pi}(q^{2}) , 
\end{eqnarray}

\begin{eqnarray}
<\pi^{-}(p_{1})\pi^{+}(p_{2})f_{0}(p_{3}) \vert (\bar{d}d) \vert 0> =v\dfrac{f_{B}m_{B}^{2}}{f_{\pi}m_{b}}(1-\dfrac{s_{13}-m_{1}^{2}-m_{3}^{2}}{m_{B}^{2}-m_{3}^{2}}) F^{\phi\pi\pi}(q^{2}) , 
\end{eqnarray}
where 
\begin{eqnarray}
v=\dfrac{m_{\pi}^{2}}{m_{u}+m_{d}}.
\end{eqnarray}
Form factor $F(q^2)$ in Eq. (13) is described as

\begin{equation}
F^{M_{1}M_{2}M_{3}}(q^{2}) = \dfrac{1}{1-q^{2}/\Lambda^2_{x}} .
\end{equation}

We intend to compute the branching ratios for $ B\longrightarrow \phi \pi \pi $ decay in the improved QCD factorization approach. In order to this purpose, considering the vertex corrections to this decay indicated as $ f_{I} $ and$ f  _{II}  $  in factorization. Their effects can be mixed with the Wilson coefficients~\cite{Pobylitsa:2001gmt} \
\begin{align}
\nonumber
&a_{2}=c_{2}+\dfrac{c_{1}}{3}+\dfrac{\alpha_{s}}{4\pi}\dfrac{C_{F}}{3}c_{1}(-18+12ln\dfrac{m_{c}}{\mu}+f_{I}+f_{II}),\\
\nonumber
&a_{3}=c_{3}+\dfrac{c_{4}}{3}+\dfrac{\alpha_{s}}{4\pi}\dfrac{C_{F}}{3}c_{4}(-18+12ln\dfrac{m_{c}}{\mu}+f_{I}+f_{II}),\\
\nonumber
&a_{5}=c_{5}+\dfrac{c_{6}}{3}+\dfrac{\alpha_{s}}{4\pi}\dfrac{C_{F}}{3}c_{6}(+6-12ln\dfrac{m_{c}}{\mu}-f_{I}-f_{II}),\\
\nonumber
&a_{7}=c_{7}+\dfrac{c_{8}}{3}+\dfrac{\alpha_{s}}{4\pi}\dfrac{C_{F}}{3}c_{8}(+6-12ln\dfrac{m_{c}}{\mu}-f_{I}-f_{II}),\\
&a_{9}=c_{9}+\dfrac{c_{10}}{3}+\dfrac{\alpha_{s}}{4\pi}\dfrac{C_{F}}{3}c_{10}(-18+12ln\dfrac{m_{c}}{\mu}+f_{I}+f_{II}),
\end{align}

where the term $ f_{I} $, hard scattering function, results from the vertex corrections and $ f_{II} $ is expected with hard gluon exchange involving the spectator quark in the D meson. The vertex corrections are given by~\cite{Pobylitsa:2003ju} \ 

\begin{eqnarray}
\nonumber
f_{I}=\dfrac{2\sqrt{6}}{f_{\phi}}\int{dx \phi^{L}_{\phi}(x)}[\dfrac{3(1-2x)}{1-x}ln(x)
-3\pi i+3ln(1-r^{2})+\dfrac{2r^{2}(1-x)}{1-r^{2}x}],\\
f_{II}=\dfrac{4\pi^2}{N}\dfrac{if_Rf_Bf_{\phi}}{X_s}\int _0^1{d\rho \dfrac{\phi^B_{\vert}(\rho)}{\rho}}\int _0^1{du\dfrac{\phi^{\phi}_{\vert\vert}(u)}{u}}\int _0^1{d\eta \dfrac{\phi^R_{\vert\vert}(\eta)}{\eta}}.
\end{eqnarray}
The leading twist distribution amplitudes are given in terms of
an expansion in Gegenbauer polynomials~\cite{Lu:2007sg}\
\begin{eqnarray}
\phi_{i}(u,\mu)=6u\bar{u}[1+\sum_{n=1}^{\infty}\alpha_{n,i}(\mu)C_{n}^{3/2}(2u-1)], \quad\quad  i=\bot,\| .
\end{eqnarray}

The other three-body decay that is investigated is $ B^{0}_s \longrightarrow \phi \pi^{+}\pi^{+}$ decay. Feynman diagrams for this decay can be plotted in Fig. 1. The amplitude of this decay includes $<B^{0}_s \longrightarrow \phi>\times<0\longrightarrow \pi^{+}\pi^{-}>$ and $<B^{0}_s \longrightarrow 0>\times<0\longrightarrow \phi \pi^{+}\pi^{-}>$. Thus, under the factorization hypothesis, the decay amplitude $B^{0}_s \longrightarrow \phi \pi^{+}\pi^{-}$ is given by
 \begin{align}
  \nonumber
  <  \phi \pi^{+}\pi^{-}\vert H_{eff} \vert B^{0}_s>&=\dfrac{iG_{F}}{2\sqrt{2}}((a_{2}V_{ub}V^{*}_{us}+(a_3+a_5+a_7)\lambda_p)<\phi\vert (\bar{b}s)_{V-A} \vert B^{0}_s>\\
  \nonumber
  & \times <\pi^{+}\pi^{-}\vert (\bar{u}u)_{V-A} \vert 0>+((a_3+a_5-1/2(a_7+a_9))\lambda_p\\
  \nonumber
 & \times <\phi\vert (\bar{b}s)_{V-A} \vert B^{0}_s> <\pi^{+}\pi^{-}\vert (\bar{d}d)_{V-A} \vert 0>+(a_{2}V_{ub}V^{*}_{us}\\
  \nonumber
  & +(a_3+a_5+a_7)\lambda_p)<0\vert (\bar{b}s)_{V-A} \vert B^{0}_s> <\phi \pi^{+}\pi^{-}\vert (\bar{u}u)_{V-A} \vert 0>\\
  \nonumber
& +(a_3+a_5+a_7+a_9)\lambda_p <0\vert (\bar{b}s)_{V-A} \vert B^{0}_s>\\
& \times <\phi \pi^{+}\pi^{-}\vert (\bar{d}d)_{V-A} \vert 0>.
&\end{align}\\
The hadronic matrix elements for $ B^{0}_s \longrightarrow \phi$ can be described following as~\cite{paracha:2015bm}
\begin{align}
\nonumber
<\phi (p_{3},\varepsilon)\vert (\bar{b}s)_{V-A} \vert  B^{0}_s(p_{B_s})>=i((m_3+m_{B_s})\varepsilon_\mu A_1^{B_s\phi}(q^2)\\
\nonumber
-\dfrac{\varepsilon.p_{B_s}}{m_3+m_{B^*}}(p_{B_s}+p_3)_{\mu} A_2^{B_s\phi}(q^2)\quad\quad\\
 -2m_{3}\dfrac{\varepsilon.p_{B_s}}{q^2}q_\mu (A_3^{B_s\phi}(q^2)-A_0^{B_s\phi}(q^2)),
\end{align}

where 
\begin{equation}
A_3^{B_s\phi}(q^2)=\dfrac{m_{B_s}+m_3}{2m_{3}}A_1^{B_s\phi}(q^2)-\dfrac{m_{B_s}-m_3}{2m_{3}}A_2^{B_s\phi}(q^2).
\end{equation}

The two pion creation matrix element of the weak interaction current can be expressed by
\begin{eqnarray}
<\pi^{+}(p_2)\pi^{-}(p_1)\vert (\bar{q}q)_{V-A} \vert 0>=(p_1-p_2)F_1^{\pi\pi}(q^2).
\end{eqnarray}

The nonresonant weak and electromagnetic form factor $ F^{\pi\pi} $ is parametrized as follows~\cite{Deshpande:1995nu}
\begin{eqnarray}
\nonumber
F^{\pi\pi}_{em}(q^{2})=\dfrac{1}{1-q^{2}/M_{*}^{2}+i\Gamma_{*}/M_{*}},
\end{eqnarray}
\begin{eqnarray}
F^{\pi\pi}_{weak}(q^{2})=\dfrac{F^{\pi\pi}(0)}{1-q^{2}/\Lambda_{\chi}^{2}+i\Gamma_{*}/\Lambda_{\chi}},
\end{eqnarray}
using $\Gamma_{*}=200 $ Mev, $M_{*}=600 $ Mev and $\Lambda_{\chi}=830 $ Mev is the chiral-symmetry breaking scale. By multiplying matrix element, we have
\begin{align}
\nonumber
<\phi (p_{3},\varepsilon)\vert (\bar{b}s)_{V-A} \vert  B^{0}_s(p_{B_s})><\pi^{+}(p_2)\pi^{-}(p_1)\vert (\bar{q}q)_{V-A} \vert 0>\\
\nonumber
=iF_1^{\pi\pi}(q^2)((m_{B_s}+m_3)\varepsilon.(p_1-p_2) A_1^{B_s\phi}(q^2)\\
 -2\dfrac{\varepsilon.p_{B_s}}{m_3+m_{B^*}}(p_1.p_3-p_2.p_3)A_2^{B_s\phi}(q^2)).\quad\quad
\end{align}

  \subsection{Resonant contribution} As noticed before, Dalitz plot model can indicate presence intermediate resonance. The decay of amplitude of B meson into $ \pi\pi$ in the different wave is to be appropriate to the pion non-strange scalar or vector form factor depending on the wave studied  which the three-body matrix element $ <\pi^+\pi^- \vert V_{\mu} \vert B^0>$ can be described by S, P and D-waves from $\pi\pi$ channel. Resonant effects are explained in terms of Breit-Wigner  formalism. Thus, the resonant contribution of $B^0 \rightarrow \phi \pi^+ \pi^-$ can be written as follow
 
 \begin{eqnarray}
\nonumber
<\pi^+(p_{2})\pi^-( p_{1}) \vert (\bar{b}d)_{V-A}\vert B^{0}(p_B) >^{R} =  \sum_{i}\dfrac{g^{T_i \rightarrow \pi^{+} \pi^{-}}}{s-m_{T_i}^{2}+im_{T_i}\Gamma_{{T_i}}}  \varepsilon_{\sigma\gamma}p_{1}^{\sigma}p_{1}^{\gamma} \\ 
\nonumber
\times <T\vert (\bar{b}d)_{V-A}\vert  B^{0} >-\sum_{i}\dfrac{g^{V_i \rightarrow \pi^{+} \pi^{-}}}{s-m_{V_i}^{2}+im_{V_i}\Gamma_{{V_i}}}\varepsilon^*.(p_1-p_2)\quad\quad\quad\quad\quad\\
\times<V \vert (\bar{b}d)_{V-A}\vert  B^{0} > -\sum_{i} \dfrac{g^{S_i \rightarrow \pi^{+} \pi^{-}}}{s-m_{S_i}^{2}+ im_{S_i}\Gamma_{{S_i}}}<S \vert (\bar{b}d)_{V-A}\vert  B^{0}>. \quad \quad 
  \end{eqnarray}
  
The three-body matrix element can receive contribution from $ f_2 $ tensor meson, $\rho$ vector meson and $ f_{0} $ scalar resonances. The form factor for $B \longrightarrow S\cite{Cheng:2010vk} $, $B \longrightarrow V\cite{paracha:2015bm} $ and $B \longrightarrow T$\cite{Wang:2010ni} transition are described by 
 \begin{eqnarray}
 \nonumber
< f_{0}(p_{q})\vert (\bar{b}d)_{V-A}\vert ِB^{0}(p_{B})>=-i[((p_{B}+p_{q})_{\mu}-\dfrac{m_{B}^{2}-m_{f_{0}}^{2}}{q^{2}}q_{\mu})F_{1}^{Bf_{0}}(q^{2})\\
 \nonumber
+\dfrac{m_{B}^{2}-m_{f_{0}}^{2}}{q^{2}}q_{\mu}F_{0}^{Bf_{0}}(q^{2})],\quad\quad\quad\quad\quad\quad\quad\\
  \nonumber
< \rho(p_{q},\varepsilon)\vert (\bar{b}d)_{V-A}\vert ِB^{0}(p_{B})>=i[(m_B+m_{\rho})\varepsilon_\mu^*A_1^{B\rho}(q^2)-\dfrac{\varepsilon^*.p_B}{m_B+m_{\rho}}\\
 \nonumber
 (p_B+p_q)_{\mu}A_2^{B\rho}(q^2)-2m_{\rho}\dfrac{\varepsilon^*.p_B}{q^2}q_{\mu}(A_3^{B\rho}(q^2)-A_0^{B\rho}(q^2))],\\
  \nonumber
< f_{2}(p_{q},\varepsilon)\vert (\bar{b}d)_{V-A}\vert ِB^{+}(p_{B})>=ih\varepsilon_{\mu\nu\lambda\rho}\varepsilon^{*\nu\alpha}p_{B\alpha}(p_B+p_q)^{\lambda}(p_B-p_q)^{\rho}\quad\\
+k(q^2)\varepsilon^{*\mu\nu}p_{B\nu}+\varepsilon^*_{\alpha\beta}p_B^\alpha p_B^\beta (b_+(q^2)(p_B+p_q)^{\mu}+b_-(q^2)(p_B-p_q)^{\mu}).
\end{eqnarray}

   The polarization tensor $ \varepsilon^{\mu\nu}(p_q,\lambda) $ with the momentum p and helicity $ \lambda $ is given by~\cite{Morales:2016pcq} 
  \begin{eqnarray}
\nonumber
&\varepsilon^{\mu\nu}(\pm2)=\varepsilon^{\mu}(\pm1)\varepsilon^{\nu}(\pm1) ,\quad\quad\quad\quad\quad\quad\quad\quad\quad\quad\quad\quad\quad\quad\quad\quad\quad\\
\nonumber
&\varepsilon^{\mu\nu}(\pm1)=\dfrac{1}{\sqrt{2}}(\varepsilon^{\mu}(\pm1)\varepsilon^{\nu}(0)+\varepsilon^{\mu}(0)\varepsilon^{\nu}(\pm1)) ,\quad\quad\quad\quad\quad\quad\quad\quad\quad\\
& \varepsilon^{\mu\nu }(0)=\sqrt{\dfrac{1}{6}}(\varepsilon^{\mu}(+1)\varepsilon^{\nu}(-1)+\varepsilon^{\mu}(-1)\varepsilon^{\nu}(+1))+\sqrt{\dfrac{2}{3}}(\varepsilon^{\mu}(0)\varepsilon^{\nu}(0)) ,
\end{eqnarray}
where $ \varepsilon^{\mu}(0,\pm1) $ presenting the polarization vector of massive vector state moving along the Z-axis (see Eq. (7)). We have used the partial widthes for determining coupling constant $ f_0\rightarrow \pi\pi $, $ \rho \rightarrow \pi\pi $ and $ f_2\rightarrow \pi\pi $, as~\cite{Patrignant:2016cp}
\begin{align} 
  \nonumber
\Gamma&(f_0(980) \rightarrow \pi^+\pi^-)=34.2 ^{+13.9}_{-11.8}MeV,\\ \quad\quad\quad\quad
 \nonumber
\Gamma&(f_0(1370) \rightarrow \pi^+\pi^-)=10.8 \pm 2MeV,\\
 \nonumber
\Gamma&(f_0(1500) \rightarrow \pi^+\pi^-)=35.8 \pm 4 MeV,\\
\nonumber
\Gamma&(\rho(770)\rightarrow \pi^+\pi^-)\sim 149.1 \pm 0.8 MeV,\\ 
\nonumber
\Gamma&(\rho(1450)\rightarrow \pi^+\pi^-)=400 \pm 60 MeV,\\ \quad\quad\quad\quad
\Gamma&(f_2(1270)\rightarrow \pi^+\pi^-)=165 \pm 9 MeV.\quad\quad\quad\quad
\end{align}
Also note that $g^{S\rightarrow M_1M_2}$, $g^{V\rightarrow M_1M_2}$ and $g^{T\rightarrow M_1M_2}$ are the coupling constants for scalar, vector and tensor mesons.
\begin{eqnarray}
  \nonumber
\Gamma_{S\rightarrow M_1M_2}=\dfrac{p_c}{8\pi m_S^2}g^2_{S\rightarrow M_1M_2},\quad& \Gamma_{V\rightarrow M_1M_2}=\dfrac{p_c^3}{6\pi m_V^2}g^2_{V\rightarrow M_1M_2}, \\
\quad\quad \Gamma_{T\rightarrow M_1M_2}=\dfrac{p_c^5}{15\pi m_T^4}g^2_{T\rightarrow M_1M_2},&
\end{eqnarray}
 where $p_c$ is the center of mass momentum. Thus, the resonant amplitude can be obtained as follow
\begin{align}
  \nonumber
  M_{R}(B^- \longrightarrow \pi^-(p_1) &\pi^{+}(p_2)\phi(p_3,\varepsilon))=\dfrac{iG_{F}}{\sqrt{2}}\dfrac{1}{2}(a_3+a_5+a_7) \lambda_p m_{\phi}f_{\phi}\\
  \nonumber
 & \times [ \dfrac{g^{f_2\rightarrow \pi^{+} \pi^{-}}}{s-m_{f_2}^{2}+im_{f_2}\Gamma_{f_2}}  \varepsilon_{T}^{\alpha\beta}\varepsilon_{\mu}(p_B+p_{f_2})_{\rho} \\ 
\nonumber
& \times (k(m_{\phi}^2)\delta_{\alpha}^{\mu}\delta_{\beta}^{\rho}+b_+(m_{\phi}^2)p_{3\alpha}p_{3\beta}g^{\mu\rho})\\
  \nonumber 
 & -\sum_{i}\dfrac{ig^{\rho_i \rightarrow \pi^{+} \pi^{-}}}{s-m_{\rho_i}^{2}+im_{\rho_i}\Gamma_{\rho_i}}(-(m_B+m_{\rho_i})\varepsilon.(p_1-p_2)\\
  \nonumber
 & \times A_1^{B\rho_i}(m_{\phi}^2)-\dfrac{A_2^{B\rho_i}(m_{\phi}^2)}{m_B+m_{\rho_i}}\varepsilon.(p_B+p_{\rho_i})p_B.(p_1-p_2))\\
 \nonumber
 &-\sum_{i} \dfrac{ig^{f_{0 i} \rightarrow \pi^{+} \pi^{-}}}{s-m_{f_{0 i}}^{2}+ im_{f_{0 i}}\Gamma_{f_{0 i}}}\varepsilon.(p_B+p_{f_{0i}}) F_1^{Bf_{0i}}(m_{\phi}^2)],\\
 &  \end{align}
where, the polarization tensor $ \varepsilon^{\alpha\beta}_{T}$ follows the Eq. (28). In $ B_s^0 \rightarrow \phi  \pi\pi$, the two-body matrix element $<\pi^{+}\pi^{-}\vert V_{\mu}\vert 0>$ can also receive contribution from $ f_2 $ tensor meson, $\rho$ vector meson and $ f_{0} $ scalar resonances in the $ \pi^{-}\pi^{+}$ channel. As noted resonant contribution of $ B_s^0 \rightarrow \phi  \pi\pi$ is investigated by using Breit-Wigner formalism.
  \begin{eqnarray}
\nonumber
<\pi^-( p_{1})\pi^+(p_{2}) \vert (\bar{q}q)_{V-A}\vert 0 >^{R} = \sum_{i} \dfrac{g^{T_i \rightarrow \pi^{+}\pi^{-}}}{s-m_{T_i}^{2}+im_{T_i}\Gamma_{{T_i}}}  \varepsilon_{\sigma\gamma}p_{1}^{\sigma}p_{1}^{\gamma} \\ 
\nonumber
\times <T_i\vert (\bar{q}q)_{V-A}\vert 0>-\sum_{i}\dfrac{g^{V_i \rightarrow \pi^{+} \pi^{-}}}{s-m_{V_i}^{2}+im_{V_i}\Gamma_{{V_i}}}\varepsilon^*.(p_1-p_2)\quad\quad\quad\\
\times<V_i \vert (\bar{q}q)_{V-A}\vert  0 > - \sum_{i}\dfrac{g^{S_i \rightarrow \pi^{+} \pi^{-}}}{s-m_{S_i}^{2}+ im_{S_i}\Gamma_{{S_i}}}<S_i \vert (\bar{q}q)_{V-A}\vert  0>. 
  \end{eqnarray}
  
   The decay constants of scalar and vector meson are defined~\cite{Zhang:2016qvq, Chen:2011ut}
\begin{eqnarray}
\nonumber
<0 \vert (\bar{q}q)_{V-A} \vert f_0(980)(p_{f_{0}})>=f_{f_0}p_{f_0}^\mu,\\
<0 \vert (\bar{q}q)_{V-A} \vert \rho(p_\rho,\varepsilon)>=f_{\rho}m_{\rho}\varepsilon_{\mu}^*.
\end{eqnarray}
  
 The polarization of tensor meson satisfies the following relation~\cite{Datta:2007yk}
 \begin{eqnarray}
\varepsilon^{\mu\nu}=\varepsilon^{\nu\mu},\quad\quad \varepsilon_\mu^\mu=0,\quad\quad p_\mu\varepsilon^{\mu\nu}=p_\nu\varepsilon^{\mu\nu}=0.
\end{eqnarray}
Therefore, the decay constant of tensor meson is defined 
\begin{eqnarray}
<0 \vert (\bar{q}q)_{V-A} \vert ِT(p_{q},\varepsilon)>=a\varepsilon_{\mu\nu}p^\nu+b\varepsilon_{\mu\nu}p^\mu=0.
\end{eqnarray}

 Then we are led to
 
 \begin{align}
 \nonumber
<\phi(&p_{3},\varepsilon) \vert (\bar{b}s)_{V-A}\vert B_s^0( p_{B_s}) ><\pi^+(p_{2})\pi^-( p_{1}) \vert (\bar{q}q)_{V-A}\vert 0 >^{R} \\
\nonumber
&=-\sum_i i\dfrac{g^{f_{0i} \rightarrow \pi^{+}\pi^{-}}}{s-m_{f_{0i}}^{2}+ im_{f_{0i}}\Gamma_{{f_{0i}}}}
f_{f_{0i}}((m_{B_s}+m_3)\varepsilon.p_{B_s} \\
\nonumber
& \times A_1^{B_s\phi}(q^2)-\dfrac{\varepsilon.p_{B_s}}{m_{B_s}+m_{\phi}}(p_{B_s}+p_3).p_{f_{0i}}A_2^{B_s\phi}(q^2)\\
\nonumber
& - 2m_3\dfrac{\varepsilon.p_{B_s}}{q^2}q.p_{f_{0i}}(A_3^{B_s\phi}(q^2)-A_0^{B_s\phi}(q^2))\\
\nonumber
& -\sum_i\dfrac{g^{\rho_i \rightarrow \pi^{+} \pi^{-}}}{s-m_{\rho_i}^{2}+ im_{\rho_i}\Gamma_{{\rho_i}}}\varepsilon_{\rho_i}.(p_1-p_2)f_{\rho_i}m_{\rho_i}((m_{B_s}+m_3)\varepsilon.\varepsilon_{\rho_i}\\
\nonumber
&\times A_1^{B_s\phi}(q^2)-\dfrac{\varepsilon.p_{B_s}}{m_{B_s}+m_{\phi}}\varepsilon_{\rho_i}.(p_{B_s}+p_3)A_2^{B_s\phi}(q^2)\\
& - 2m_3\dfrac{\varepsilon.p_{B_s}}{q^2}\varepsilon_{\rho_i}.q(A_3^{B_s\phi}(q^2)-A_0^{B_s\phi}(q^2)).
 \end{align}
 
 where $q=p_1+p_2=p_{B_s}-p_3$ and $ p_{R}=p_1+p_2$. Therefore 
  \begin{eqnarray}
 \varepsilon.p_{f_0}= \varepsilon.p_{B_s}=\dfrac{m_{B_s}}{m_3}\vert p_3 \vert.
   \end{eqnarray}
    Finally the decay amplitude through resonance intermediate reads
    \begin{align}
    \nonumber
   M_R(B_s^0 \rightarrow \phi \pi^+ \pi^-)&=-\sum_i i\dfrac{g^{f_{0i} \rightarrow \pi^{+}\pi^{-}}}{s-m_{f_{0i}}^{2}+ im_{f_{0i}}\Gamma_{{f_{0i}}}} f_{f_{0i}}  m_{B_s}\vert p_3 \vert\\
   \nonumber
& \times 2A_0^{B_s\phi}(s_{12}) -\sum_i\dfrac{g^{\rho_i \rightarrow \pi^{+} \pi^{-}}}{s-m_{\rho}^{2}+ im_{\rho}\Gamma_{{\rho}}}f_{\rho_i}m_{\rho_i}\\
\nonumber
& \times [-(m_{B_s}+m_3)\varepsilon.(p_1-p_2)A_1^{B_s\phi}(s_{12})\\
&-\dfrac{\varepsilon.p_{B_s}}{m_{B_s}+m_3} (2p_3.p_2-2p_3.p_1)A_2^{B_s\phi}(s_{12})].
  \end{align}
The decay width of three-body is written by~\cite{Cheng:2002qu}
\begin{eqnarray}
\nonumber
\Gamma(B\longrightarrow M_{1}M_{2}M_{3})=\dfrac{1}{(2\pi)^{3}32M_{B}^{2}} \int_{s_{min}}^{s_{max}} \int_{t_{min}}^{t_{max}}\vert M_{NR}(B\longrightarrow M_{1}M_{2}M_{3})\quad\quad\\
\quad\quad\quad\quad\quad +M_{R}(B\longrightarrow M_{1}M_{2}M_{3}) \vert ^{2} dtds,
\end{eqnarray}
where,
\begin{eqnarray}
\nonumber
 s_{min}&=&(m_{1}+m_{2})^{2} ,\\
\nonumber
 s_{max}&=&(m_{B}-m_{3})^{2} ,\\
\nonumber
t_{min,max}&=&m^2_{2}+m^2_{3}-\dfrac{1}{s}[(s-m^2_{B}+m^2_{1})\times(s+m^2_{2}-m^2_{3})\\
&\pm & \lambda^{{1}/{2}}(s, m_{B}^{2}, m_{1}^{2})\lambda^{{1}/{2}}(s, m_{2}^{2}, m_{3}^{2}),
\end{eqnarray}
where, $\lambda(x, y, z)=x^2+y^2+z^2-2(xy+xz+yz)$.\\

     \section{Numerical Result}

We require to determine the input of several physical ingredients for numerical analysis. The Fermi coupling constant, $G_{F}$, is taken to be equal to $1.66\times10^{-5} GeV$. The value of the decay constants and meson masses in units of MeV are~\cite{ Patrignant:2016cp,Zhang:2016qvq, Zhang:2010km} \
\begin{align}
\nonumber
m_{\pi^{\pm}}=139.57\pm 0.00035 ,  m_{\phi}=1019\pm 0.016,  m_{B^0}=5279.63 \pm 0.15, \quad\quad\quad\\
\nonumber
m_{f_{0}}=990\pm 20, m_{B_s}=5366.89 \pm 0.19 ,m_{\rho}=775.26\pm 0.25, \quad\quad\quad\quad\quad\quad\\
\nonumber
m_{f_2}=1275.5\pm 0.8, m_{B*}= 5324.65\pm 0.25, f_{f_0}=0.37,f_{\phi}=221, \quad\quad\quad\quad\\
\nonumber
 f_{\pi}=130, f_{\rho}=218\pm 2, f_{B}=176 \pm 42, f_{B^*}=194 \pm 6. \quad\quad\quad\quad\quad\quad\quad\quad
\end{align}

\begin{table}[h!]
\centering
\caption{The value of wilson coefficients $c_{i}$ at three renormalization scale $\mu$.}
\begin{tabular}{|c| c| c| c| c |c |c |c |c |c |c |c|}
\hline   $\mu$  &    $c_{1}$  &$c_{2}$ & $c_{3}$ & $c_{4}$ & $c_{5}$ & $c_{6}$ & $c_{7}/\alpha $  & $c_{8}/\alpha $& $c_{9}/\alpha$& $c_{10}/\alpha $  \\  \hline
 $m_{b}/2$   & 1.137  & -0.295 & 0.021 & -0.051 & 0.010 & -0.065 &-0.024 & 0.096 & -1.325 & 0.331 \\  \hline
  $m_{b}$   & 1.081  & -0.19 & 0.014 & -0.036 & 0.009 & -0.042 &-0.011 & 0.060 & -1.254 & 0.223\\  \hline
   $2m_{b}$   & 1.045  & -0.113 & 0.009 & -0.025 & 0.007 & -0.027 &-0.011 & 0.039 & -1.195 & 0.144 \\  \hline
\end{tabular}
\end{table}

The coupling constants are $g_{B^*B\pi}=32\pm5$ ~\cite{Melikhov:2000yu} and $g_{B_s^*BK}=10.6$ ~\cite{Cerqueira:2011za}. The value of $C_{i}$ at three scales $\mu=m_{b}/2$, $ m_{b} $, $ 2m_{b} $ is given in Table 1 ~\cite{Beneke:2001ev}. The form factors $A_i^{B\rho}$ are depicted in Table 2 ~\cite{Wu:2006rd}. The form factors of transition $ B_s \rightarrow \phi$ at $ q^2=0 $ applied in Eq. (20) are given in Table 3 ~\cite{Paracha:2015mrd}. The $ B \rightarrow f_2$ transition form factor is depicted in Table 4 ~\cite{Cheng:2010yd}. The parameters in the form factor of $ B \rightarrow f_0(d\bar{d})$ transition is $0.3\pm 0.05$ ~\cite{Colangelo:2010bg}. Branching ratios of these decays at three scales are calculated as shown on the Table 5.
 
\begin{table}[h!]
\centering
\caption{The form factor for $ B\rightarrow \rho$ transition.}
\begin{tabular}{|c| c| c| c| c|}
\hline  decay  &     \multicolumn{2}{|c|}{F(0) }   &     $ a_{F} $    &      $ b_{F} $   \\  \hline
    &$  A_1 $ &  0.232  & 0.42  & -0.25 \\ 
     $  B\rightarrow \rho$  &   $ A_2 $ &  0.187 & 0.98 & -0.03 \\  
      &$  A_3 $ &  -0.221  & 1.16  & 0.09 \\ 
       &$  V $ &  0.289  & 1.32  & 0.34 \\ \hline
    
\end{tabular}
\end{table}

\begin{table}[h!]
\centering
\caption{The parameters $ B_s\rightarrow \phi $ transition.}
\begin{tabular}{|c| c| c| c| c|}
\hline  decay  &   $ A_1$  &    $  A_2$    &      V   \\  \hline
        $ B_s\rightarrow \phi $   &  0.29  & 0.25  & 0.24\\ \hline
\end{tabular}
\end{table}

\begin{table}[h!]
\centering
\caption{The parameters for $ B\rightarrow  f_2$  transition.}
\begin{tabular}{|c| c| c| c| c|}
\hline  decay  &     k &     $ b_+ $    &      $ b_- $   \\  \hline
     $  B\rightarrow f_2  $   &   0.425    & -0.014 & 0.014 \\  \hline
  
\end{tabular}
\end{table}

\begin{table}[h!]
\centering
\caption{The branching ratio of decays  at scales $\mu=m_{b}/2$, $ m_{b} $, $ 2m_{b} $.}
\begin{tabular}{|c| c| c| c| c| c| }
\hline  mode   & $B^0 \rightarrow \phi \pi^{+}\pi^{-}$ 
 & $B^0_s\rightarrow \phi \pi^{+}\pi^{-}$      \\ 
 & ($\times 10^{-7}$)& ($\times 10^{-6}$) \\ \hline
BR$(2m_b)$  & $ 2.16 \pm 0.21 $ & $5.5 \pm 0.20 $   \\  \hline
  BR$(m_b)$& $ 1.69 \pm 0.19 $ & $3.28 \pm 0.17 $   \\  \hline
 BR$(m_b/2)$ & $ 0.98 \pm 0.15$ & $2.34 \pm0.13 $   \\  \hline
BR$_{\textrm{Exp}}$ & $1.82 \pm 0.25 $& $3.48 \pm 0.23  $  \\  \hline
\end{tabular}
\end{table}

\section{Conclusion}

In this research, we have computed the branching ratios of $B^{0}\rightarrow \phi \pi^{+}\pi^{-}$ and $ B^0_s\rightarrow \phi \pi^+\pi^- $ decays obtained from studies of three-body decays. The branching ratios of hadronic three-body decays are evaluated by applying the factorization approach. The Dalitz plot model for  decays is determined by considering several resonant and nonresonant amplitudes. In order to examine the branching ratio of three-body decays, we investigated resonant and nonresonant contributions. There are factorizable and non-factorizable contributions that the non-factorizable terms corresponding to the hard spectator interactions and vertex corrections are computed. That's why, the improved QCD factorization approach was applied. Eventually, we computed the branching ratios at three scales $ m_b/2 $, $ m_b $ and $ 2m_b $.
In generalized factorization, the computed branching ratio of $B^{0}\rightarrow \phi \pi^{+}\pi^{-}$ and $ B^0_s\rightarrow \phi \pi^+\pi^- $  the value  $ 1.69 \pm 0.19$ and $3.28 \pm 0.17$ at scale $m_b$ while the experimental result $1.82 \pm 0.25 $ and $3.48 \pm 0.23  $ respectively. The comparison between our obtained value and experiment indicates in relative agreement with experimental information which the value of branching ratio on the scale $ m_b $ is correspondent with the experimental values. The resonant contribution which considered in computation of branching ratio in $ B^0_s\rightarrow \phi \pi^+\pi^- $  decay is dominant and it's approximately correspondent with numerical values from PDG arranged. Also, in $ B^0\rightarrow \phi \pi^+\pi^- $  decay, the non-resonant contribution which considered in our computation is dominant and it's approximately correspondent with experimental results. In summary, calculated branching ratios for the sum of non-resonant and resonant amplitudes are consistent with experimental results.

\end{document}